\documentclass{article}
\usepackage{amsmath}
\usepackage{amsfonts}
\begin{document}
%%%------------------------------------------------------

\title{Landau electron in a rotating environment: \\ a general factorization of time evolution}

\author{J. Chee\\\\Department of Physics, School of Science,\\Tianjin Polytechnic University,
 Tianjin 300387, China}

\date{July, 2012}
\maketitle
\begin{abstract}
For the Landau problem with a rotating magnetic field and a potential in the (changing) direction of the field, we derive a general factorization of the time evolution operator that includes the adiabatic factorization as a special case. We assume that the direction of the magnetic field changes with time in a general way, so the Heisenberg equations of motion cannot be solved by quadrature. Also, the potential is assumed to be of a general form. We use the rotation operator associated with the solid angle Berry phase to transform the problem to a rotating reference frame that follows the direction of the magnetic field. In the rotating
reference frame, we derive a natural factorization of the time evolution operator by recognizing the crucial role played by a gauge transformation. The
major complexity of the problem arises from the coupling between motion in the direction of the magnetic field and motion perpendicular to the field. In the factorization, this complexity is consolidated into a single operator that approaches the identity operator when the potential confines the particle sufficiently close to a plane perpendicular to the magnetic field. The structure of this operator is clarified by deriving an expression for its generating Hamiltonian. The magnetic translation is the most notable physical consequence in the adiabatic limit. This and the non-adiabatic effects are studied as consequences of the general factorization.
\end{abstract}
\pagebreak

%%%-----------------------------------------------------

\section{Introduction}
The study of time-dependent quantum systems has intimate connections with the geometric phase concept \cite{berry, shapere,bohm} which has many applications
in physics. When the Hamiltonian is time-dependent, the time evolution operator is often nontrivial, i.e.,  $U(t)\neq\exp(-i\int_0^tH(\tau)d\tau)$, and clarifying its structure is of significance in understanding the dynamics of the system.

One may study $U(t)$ corresponding to a time-dependent Hamiltonian
by factorizing it into several operators, each of them is simpler at least in some respects than $U(t)$ itself.
A well-known example of this is found in the proof of the quantum adiabatic theorem, as presented in standard texts such as Messiah \cite{messiah}.
There the time evolution of a system in a changing environment is constructed as the product of three operators, $U=GDU_\epsilon$, where $G$ is a path dependent geometric operator that brings an initial eigenstate to an instantaneous eigenstate of the Hamiltonian, $D$ is a dynamical operator that only contribute dynamical phase factors to these eigenstates and $U_\epsilon$ approaches the identity operator in the adiabatic limit $\epsilon\rightarrow0$. The parameter $\epsilon$ determines how fast the Hamiltonian $H[{\bf R}(s)]=H[{\bf R}(\epsilon t)]$ changes with time $t$, for a given map from $(s\in)[0,1]$ to a path in the parameter space $M$ to which ${\bf R}$ belongs. For any such fixed map from $[0,1]$ to $M$, $1/\epsilon$ provides a time scale which is the total time it takes for ${\bf R}(\epsilon t)$ to travel through the given path in $M$. Another relevant time scale is provided by $\hbar/E_g$, where $E_g$ is the minimum energy gap. The adiabatic limit corresponds to $(\hbar/E_g)/(1/\epsilon)\rightarrow0$, which is made use of in the proof of the adiabatic theorem \cite{messiah, avron}. The path-dependent geometric operator is (assuming the Hamiltonian has non-degenerate eigenstates for all times)
\begin{equation}
G({\bf R}(\epsilon t))=\sum_m|\psi_m({\bf R}(\epsilon t))\rangle\langle \psi_m({\bf R}(0))|,
\end{equation}
where $|\psi_m({\bf R}(\epsilon t))\rangle$ is the instantaneous eigenstate of $H[{\bf R}(\epsilon t)]$
that satisfies
\begin{equation}
\langle \psi_m({\bf R}(\epsilon t))|\dot \psi_m({\bf R}(\epsilon t))\rangle=0.
\end{equation}
The expressions for the operators $D$ and $U_\epsilon$ also involve the use of eigenstates of the Hamiltonian.

This method of factorizing $U$ using instantaneous eigenstates has certain limitations when dealing with degenerate (including infinitely degenerate) energy eigenstates. For instance, in various Landau systems involving a charged particle in time-dependent electromagnetic fields, the instantaneous energy levels can be highly degenerate. Then there is no known general method for obtaining useful information
on $G$ (which contains information on non-Abelian Berry phase) or $U_\epsilon$ using instantaneous eigenstates. However, in a specific problem where the Hamiltonian is given, one may use the algebraic structure of the Hamiltonian without referring to individual eigenstates to directly construct a factorization of $U$ that can then be applied to any representation and the associated eigenstates. From this perspective, there seems to be more problems that can be explored.

This change in perspective also allows us to seek useful factorizations of $U$ not limited by the specific form $U=GDU_\epsilon$, as long as the factorization can help clarify the structure of the total time evolution operator $U$.

\section{The problem and general considerations}
The Landau problem is of significance in many areas in physics and its variations (see, for instance, \cite{dodonov,chee,date,hashimi}) have often been discussed. In this paper our purpose is to study a
charged particle in a rotating magnetic field and a confining potential in the direction of the magnetic field. The Hamiltonian is
\begin{equation}
H=\frac{1}{2m}\big({\bf p}-e{\bf A}({\bf r},t)\big)^2 + V({\bf r}\cdot{\bf n}(\epsilon t)-L),
\end{equation}
where
\begin{equation}
{\bf A}({\bf r},t)=\frac{1}{2}B{\bf n}(\epsilon t)\times{\bf r}.
\end{equation}
The confining potential $V\big({\bf r}\cdot{\bf n}(\epsilon t)-L\big)$ is in the (changing) direction of the magnetic field $B{\bf n}(\epsilon t)$ and has
equilibrium position at the plane ${\bf r}\cdot{\bf n}(\epsilon t)-L=0$ which is perpendicular to ${\bf n}(\epsilon t)$. The distance between
this plane and the origin of the coordinate system is $L$. This can be seen as an extension of the usual Landau problem where $\bf n$ is in a fixed direction.

Here, $V\big({\bf r}\cdot{\bf n}(\epsilon t)-L\big)$ is assumed to be of a general form.
It can be a harmonic oscillator potential or other types of potentials that in general correspond to nonlinear Heisenberg equations of motion. The complexity of the problem is mainly caused by the coupling between motion in the magnetic field direction and motion perpendicular to it. The question we ask is whether the time evolution operator has a factorization that reduces to a simple form when the confinement is strong enough so that the particle can be seen as staying very close to the rotating plane. There are three time scales involved in the problem:
\begin{equation}
T_1=1/\epsilon, ~~T_2=2\pi/\omega,~~T_3=\hbar/\Delta,
\end{equation}
where $\omega=|e|B/m$ is the cyclotron frequency and $\Delta$ is the minimum energy gap determined by the Hamiltonian
\begin{equation}
\frac{1}{2m}({\bf p}\cdot{\bf n})^2+V\big({\bf r}\cdot{\bf n}(\epsilon t)-L\big).
\end{equation}
The general factorization we derive later does have a simple form when the confinement is strong in the sense that $T_3\ll T_1$ and $T_3\ll T_2$. This factorization is from a different perspective than a factorization that results in a simple behavior when $\epsilon$ is small (i.e., $T_3\ll T_1$ and $T_2\ll T_1$).
Here we point out that this different perspective seems to be more natural because, in this problem, it seems more appropriate to consider the adiabatic limit to be a special case of the strong confinement limit. In other words, the adiabatic limit here should mean
\begin{equation}
T_2/T_1\rightarrow0,~~T_3/T_2\rightarrow0.
\end{equation}
The conditions $T_2/T_1\rightarrow0$ and $T_3/T_1\rightarrow0$ are weaker than the above and do not lead to a simple behavior of the time evolution operator, since if $T_2$ is comparable to $T_3$, the coupling effect between motion in the direction of $\bf n$ and motion perpendicular to it can accumulate over the time interval $[0, T_1]$ even if $\epsilon$ is very small. This was pointed out in a special case \cite {chee} where we studied the factorization of $U$ in the adiabatic limit only and where we took $V$ to be the harmonic oscillator potential $V=\frac{1}{2}k\big({\bf r}\cdot{\bf n}(\epsilon t)-L\big)^2$. The special case, however, misses out on the more general scenario of strong confinement where $U$ already has a simple factorization. Furthermore, the adiabatic factorization contains  no information on effects due to finite confinement strength or that due to non-vanishing rotation speed. Using an improved method, in which we recognize the key role played by a gauge transformation, our purpose here is to derive a factorization of $U$ that applies for all confining potentials and an arbitrary time variation of ${\bf n}(\epsilon t)$.

The electric field ${\bf E}(t)=-\frac{\partial\bf A}{\partial t}$ corresponding to the vector potential $\frac{1}{2}B{\bf n}(\epsilon t)\times{\bf r}$ is a notable point although the potential is quite commonly used in the literature. (Here we assume that the magnetic field induced by the change of ${\bf E}(t)$ and so on are negligible.) The subtlety lies in that the vector potential is time dependent, unlike a vector potential that corresponds to a static magnetic field. Imagine we choose a different vector potential through a time dependent gauge transformation, then we have the same rotating magnetic field but a different electric field. Besides considerations based on symmetry and simplicity, there is no physical reason why we prefer this vector potential to other potentials. In reality, the induced electric field can be different if the boundary conditions that produce the rotating magnetic field are different. Though a rotating magnetic field is ubiquitous in circumstances ranging from an electric generator to a rotating neutron star \cite{bocquet}, the treatment of the associated electric field can be a complicated problem that involves moving boundary conditions. Let us note however, that $\frac{1}{2}B{\bf n}(\epsilon t)\times{\bf r}$ can be realized in a simple physical situation where the boundary is not moving at all. Consider two fixed and mutually perpendicular solenoids whose symmetry axes meet at the origin. The magnetic fields in the two solenoids have the same magnitude and change sinusoidally with $\pi /2$ phase difference. This produces a total magnetic field inside the common region that rotates about a third perpendicular axis through the origin. The induced electric field due to the change of magnetic field in each solenoid alone can be found by Faraday's law and by symmetry. The total induced electric field inside the intersection of the two solenoids can then be found to be equal to $\frac{1}{2}B\dot{\bf n}(\epsilon t)\times{\bf r}$. Adding a uniform and constant magnetic field along the third axis, we may obtain a uniform magnetic field that keeps at an arbitrary fixed angle with and rotates about the third axis whose vector potential is still given by $\frac{1}{2}B{\bf n}(\epsilon t)\times{\bf r}$.

The induced electric field does not have to be $\frac{1}{2}B\dot{\bf n}(\epsilon t)\times{\bf r}$. Imagine for instance that the cross section of one of the solenoids is not a circle but instead a square, then symmetry argument no longer applies in calculating the electric field. One still produces the same uniform time dependent magnetic field, but the induced electric field is different and it is in general nonlinear with respect to the coordinates. So the vector potential is in general nonlinear. These different vector potentials should have different physical consequences but these will not be discussed further in this paper.

\section{The rotation transformation}
The time evolution operator $U(t)$ satisfies
\begin{equation}
i{\dot U}(t)=HU(t),\  \ U(0)=I.
\end{equation}
Here, we choose units in which $\hbar=c=1$ and we shall assume that repeated indices are summed over.

Our starting point is to use the parallel transport method \cite{berry1} to transform to a rotating reference frame
that follows the direction of the magnetic field. This rotating frame has the same origin as the initial coordinate system but with a rotating basis $({\bf e}_1, {\bf e}_2, {\bf e}_3)$ which is fixed by
\begin{equation}
{\dot {\bf e}}_i(\epsilon t)=~({\bf n}\times \dot{\bf n})\times {\bf e}_i(\epsilon t),
\end{equation}
where $i=1,2,3$. The initial conditions are
\begin{equation}
{\bf e}_1(0)={\bf n'}(0),~~ {\bf e}_2(0)={\bf n}(0)\times{\bf n'}(0),~~{\bf e}_3(0)={\bf n}(0).
\end{equation}
In the above, the prime in ${\bf n'}(0)$ means the derivative with respect to arc
length $s$ at $t=0$ while ${\dot {\bf e}}_i(\epsilon t)$ denotes the time derivative. ( We assume that $ds/dt\geq 0$ for all $t$.)
It can be verified easily that ${\bf e}_3(\epsilon t)={\bf n}(\epsilon t)$ and that the three vectors form an
orthonormal basis for all $t$. Furthermore the vectors ${\bf e}_1(\epsilon t)$ and ${\bf e}_2(\epsilon t)$ satisfy the condition
$\dot{\bf e}_1(\epsilon t)\cdot{\bf e}_2(\epsilon t)=0$. Thus, when viewed as tangent vectors on the unit 2-sphere formed by all possible ${\bf n}$'s,
${\bf e}_1(\epsilon t)$ and ${\bf e}_2(\epsilon t)$ have the geometric meaning of being parallel transported along the curve ${\bf n}(\epsilon t)$ on the 2-sphere.
Now accompanying such a transformation of the coordinate system we have
\begin{equation}
U(t)=R(\epsilon t){U_1}(t)
\end{equation}
where ${U_1}$ describes the quantum time evolution in the rotating reference frame, and
\begin{equation}
R(\epsilon t)=\mathrm{T}\exp \big(-i\int_{0}^{t}({\bf n}\times \dot{\bf n})\cdot{\bf
J}~d\tau\big)=\mathrm{P}\exp \big(-i\int_{{\bf n}(0)}^{{\bf n}(\epsilon t)}({\bf n}\times d{\bf n})\cdot{\bf J}\big).
\end{equation}
$R(\epsilon t)$ is a rotation operator well-known to be associated with the ``solid angle Berry phase" \cite{berry}. It is characterized by the fundamental property that although it is only written as a time ordered exponential, it is in fact an explicit operator for a finite dimensional irreducible representation of the angular momentum
and that if $|m\rangle$ is an eigenstate of ${\bf n}(0)\cdot {\bf J}$ with ${\bf n}(0)\cdot {\bf J}|m\rangle=m|m\rangle$ and ${\bf n}(\epsilon t)$ returns to ${\bf n}(0)$ at time $T$, then $R(\epsilon T)|m\rangle=\exp(-im\Omega)|m\rangle$, where $\Omega$ is the solid angle that the closed path traversed
by $\bf n$ subtends at the origin of the parameter space. (For a review of this property, see section 2.1 of Ref \cite{rotation}.)

The operator $R$ has the property
\begin{equation}
R^{-1}(\epsilon t){\bf e}_i(\epsilon t)\cdot {\bf v}R(\epsilon t)~=~{\bf e}_i(0)\cdot {\bf v}\equiv v_i,~~i=1,2,3,
\end{equation}
where $\bf v$ is any vector operator. This property can be proven by first noting that it clearly holds for $t=0$. Now it suffices to
prove that the left hand side is constant, or its time derivative is zero. This is easily verified
by making use of the equations $\dot R= -i({\bf n}\times \dot{\bf n})\cdot{\bf J}R$, ~${\dot R}^{-1}=R^{-1}(t)i({\bf n}\times \dot{\bf n})\cdot{\bf J}$,
~${\dot {\bf e}}_i=({\bf n}\times \dot{\bf n})\times {\bf e}_i$, and the formula
$[{\bf a\cdot J}, {\bf{b\cdot v}}]=~i\bf{(a\times b)\cdot v}$ which holds for a vector operator $\bf v$.
From this property it is clear that ${U_1}^{-1}v_i{U_1}{\bf e}_i(\epsilon t)=U^{-1}v_iU{\bf e}_i(0)$. This is consistent with the fact that while the time evolution in the inertial reference frame is given by $U(t)$, seen from the rotating frame, it is described by ${U_1}$.

Using the above property, we can Taylor expand $V({\bf r}\cdot{\bf n}(\epsilon t)-L)$ and it follows that
\begin{equation}
R^{-1}(\epsilon t)V({\bf r}\cdot{\bf n}(\epsilon t)-L)R(\epsilon t)~=~V(x_3-L).
\end{equation}
Now we make the substitution $U=RU_1$ in the Schr\"{o}dinger equation for $U$. It is straightforward to obtain the following equation for ${U_1}$
\begin{equation}
i{\dot U}_1(t)=H_1{U_1}(t),\ \ {U_1}(0)=I,
\end{equation}
where $H_1$ is the Hamiltonian in the rotating frame:
\begin{equation}
H_1= \frac{1}{2m}\big({\bf p}-e{\bf A}({\bf r})\big)^2 + \alpha_2(\epsilon t)J_1-\alpha_1(\epsilon t)J_2+V(x_3-L)
\end{equation}
with
\begin{equation}
{\bf A}({\bf r})={\bf A}({\bf r},0)=B{\bf n}(0)\times {\bf r}/2,
\end{equation}
and
\begin{equation}
\alpha_{\mu}(\epsilon t)=\dot{\bf n}(\epsilon t)\cdot{\bf e}_{\mu}(\epsilon t), ~~\mu=1,2, ~~J_i=\epsilon_{ijk}x_jp_k.
\end{equation}
The magnetic field is now in a fixed direction in the rotating frame. Other rotation operators can achieve the same goal but they introduce an additional
term in $H_1$ that is proportional to $J_3$. The operator $R$ leads to the simplest form of $H_1$.

\section{General factorization of $U$}
In this section our purpose is to find a factorization of $U_1$ that has a simple form in the strong confinement limit. Suppose the confinement is strong, then motion of the particle in the direction of the magnetic field is dominated by the potential $V(x_3-L)$. Also, the wave packet should be close to the $x_3=L$ plane. However, directly extracting the operator $\exp[-i(p^2_3/2m+V(x_3-L))t]$ from $U_1$ does not lead to a useful factorization. This is because the term $\alpha_2(\epsilon t)J_1-\alpha_1(\epsilon t)J_2$ in $H_1$ contains both $x_3$ and $p_3$, and if we assign $x_3=L$ in the limit of strong confinement, then $p_3$ is completely uncertain according to the Heisenberg uncertainty principle. The issue here is that $p_3$ is not the kinematical momentum in the rotating frame in the current gauge. It is only after a gauge transformation that it is. This is the key observation that leads to the factorization of $U_1$.

First let us rewrite $H_1$ in the following form
\begin{equation}
H_1= \big(\frac{1}{2m}K_1^2+ \frac{1}{2m}K_2^2-eA_{\mu}\alpha_{\mu}x_3+V_c\big)+\big(\frac{1}{2m}K_3^2+V(x_3-L)\big)
\end{equation}
where $K_1$, $K_2$ and $K_3$ are the components of the kinematical momentum in the rotating frame, and $V_c$ is the potential for the centripetal force:
\begin{eqnarray}
K_{\mu}&=&p_{\mu}-eA_{\mu}-m\alpha_{\mu}(\epsilon t)x_3,  \   \mu=1,2, \\ K_3&=&p_3+m\alpha_1(\epsilon t)x_1+m\alpha_2(\epsilon t)x_2,\\
V_c&=&-\frac{m}{2}\dot{\bf n}^2(t)x_3^2-\frac{m}{2}\big[\alpha_1(t)x_1+\alpha_2(t)x_2\big]^2,
\end{eqnarray}
where $A_\mu, ~\mu=1,2~$ are given by $A_\mu={\bf A}({\bf r})\cdot{\bf e}_\mu$, or
\begin{equation}
A_1=-\frac{B}{2}x_2, ~A_2=\frac{B}{2}x_1.
\end{equation}
In the rotating frame, in addition to the centripetal force, there are also the Coriolis and the Euler forces. These are produced by the terms $-m\alpha_{\mu}(\epsilon t)x_3$ and $m\alpha_1(\epsilon t)x_1+m\alpha_2(\epsilon t)x_2$ in $K_{\mu}$ and $K_3$. The term $-eA_{\mu}\alpha_{\mu}x_3$ in $H_1$ is responsible for the force due to the induced electric field caused by the rotation of the magnetic field and the Lorentz force arising from the rotation of the frame $({\bf e}_i(\epsilon t), i=1,2,3)$, i.e., the charged particle has the additional velocity $({\bf n}\times \dot{\bf n})\times{\bf r}$ relative to the inertial frame $({\bf e}_i(0), i=1,2,3)$ and this has a contribution to the Lorentz force.

Our main step in elucidating the structure of $U_1$ is to perform a gauge transformation as follows. Let $|\psi(0)\rangle$ be an initial quantum state whose time evolution in the rotating frame is given by $|\psi_1(t)\rangle=U_1(t)|\psi(0)\rangle$. Then $i|{\dot \psi}_1(t)\rangle=H_1|\psi_1(t)\rangle$. In the rotating frame, we consider the gauge transformation
\begin{equation}
|\psi_1(t)\rangle=g(\epsilon t)|\psi_0(t)\rangle,
\end{equation}
where
\begin{equation}
g(\epsilon t)=\exp[-im(\alpha_1(\epsilon t)x_1+\alpha_2(\epsilon t)x_2)(x_3-2L)].
\end{equation}
In the gauge corresponding to $|\psi_0(t)\rangle$, the Schr\"{o}dinger equation is
\begin{equation}
i|{\dot \psi}_0(t)\rangle=H_0|\psi_0(t)\rangle,
\end{equation}
where
\begin{equation}
H_0=g^{-1}(\epsilon t)H_1g(t)-ig^{-1}(\epsilon t){\dot g}(\epsilon t).
\end{equation}
If the time evolution operator corresponding to $H_0$ is $U_0$, then we have
\begin{equation}
U_1=g(\epsilon t)U_0g^{-1}(0), ~~ i{\dot U_0}(t)=H_0U_0(t), ~~U_0(0)=I.
\end{equation}
Using the formula $\exp(-B)A\exp(B)=A+[A,B]$ with the condition that $[A, B]$ commutes with $A$ and $B$, we have
\begin{equation}
H_0=\frac{1}{2m}\Pi_1^2+ \frac{1}{2m}\Pi_2^2-eA_{\mu}\alpha_{\mu}x_3+V_c-ig^{-1}(\epsilon t){\dot g}(\epsilon t)+\frac{1}{2m}p_3^2+V(x_3-L)
\end{equation}
where
\begin{equation}
\Pi_\mu=p_{\mu}-eA_{\mu}-2m\alpha_{\mu}(\epsilon t)(x_3-L),  ~~\mu=1,2.
\end{equation}
This implies that in the gauge corresponding to $|\psi_0(t)\rangle$, the kinematical momentum is $(\Pi_1, \Pi_2, p_3)$.

The Hamiltonian $H_0$ corresponding to this new gauge is unique in the sense that except $\frac{1}{2m}p_3^2$, the rest of the terms in it do not depend on
$p_3$, and also $\Pi_1$ and $\Pi_2$ are in the simplest form in the limit of strong confinement. In fact, $H_0$ can be rewritten as
\begin{equation}
H_0=H_{1d}+H_{2d}+H_{\xi_0},
\end{equation}
where
\begin{equation}
H_{1d}=\frac{1}{2m}p_3^2+V(x_3-L)-\frac{m}{2}L^2{\dot{\bf n}}^2,
\end{equation}
\begin{multline}
H_{2d}=\frac{1}{2m}[(p_1-eA_1)^2+(p_2-eA_2)^2]-eL\alpha_\mu(\epsilon t)A_\mu\\+mL[{\dot \alpha}_1(\epsilon t)x_1+{\dot \alpha}_2(\epsilon t)x_2]-\frac{m}{2}[\alpha_1(\epsilon t)x_1+\alpha_2(\epsilon t)x_2]^2,
\end{multline}
\begin{equation}
H_{\xi_0}=[-2\alpha_\mu(\epsilon t)(p_\mu-eA_\mu)-eA_\mu\alpha_\mu(\epsilon t)-m{\dot \alpha}_\mu(\epsilon t)x_\mu]{\xi_0}-mL{\dot{\bf n}}^2{\xi_0}+\frac{3m}{2}{\dot{\bf n}}^2{\xi_0}^2,
\end{equation}
with
\begin{equation}
{\xi_0}= x_3-L.
\end{equation}
If we now extract the operator
\begin{equation}
U_{1d}=\exp{-i\int_{0}^{t}H_{1d}dt'}=\exp{-i[(\frac{1}{2m}p_3^2+V(x_3-L))t-\int_{0}^{t}\frac{m}{2}L^2{\dot{\bf n}}^2dt']}
\end{equation}
from the operator $U_0$, such that
\begin{equation}
U_0=U_{1d}\mathfrak{U},
\end{equation}
then $\mathfrak{U}$ satisfies
\begin{equation}
i\dot{\mathfrak{U}}=\mathfrak{H}\mathfrak{U},
\end{equation}
where $\mathfrak{H}=H_{2d}+U_{1d}^{-1}H_{\xi_0} U_{1d}$.
Define
\begin{equation}
\xi(t)=U_{1d}^{-1}(t)(x_3-L)U_{1d}(t),
\end{equation}
then $\mathfrak{H}$ can be written as
\begin{equation}
\mathfrak{H}=H_{2d}+H_\xi,
\end{equation}
where
\begin{equation}
H_\xi=[-2\alpha_\mu(p_\mu-eA_\mu)-e\alpha_{\mu}A_\mu-m{\dot \alpha}_\mu(\epsilon t)x_\mu]\xi(t)-mL{\dot{\bf n}}^2\xi(t)+\frac{3m}{2}{\dot{\bf n}}^2\xi^2(t).
\end{equation}
We see that $H_\xi\rightarrow0$ if $\xi\rightarrow0$.
Now we can write $\mathfrak{U}$ as
\begin{equation}
\mathfrak{U}=U_{2d}U_\xi
\end{equation}
where $U_{2d}$ is determined by
\begin{equation}
i\dot{U}_{2d}=H_{2d}U_{2d}, ~~U_{2d}(0)=I,
\end{equation}
and $U_\xi$ is determined by
\begin{equation}
i\dot{U}_\xi=U_{2d}^{-1}H_\xi U_{2d}U_\xi, ~~U_\xi(0)=I.
\end{equation}
Combining formulas, we obtain the factorization of the time evolution operator $U(t)$ as follows:
\begin{equation}
U(t)=R(\epsilon t)U_1(t)=R(\epsilon t)g(\epsilon t)U_{1d}(t)U_{2d}(t)U_\xi(t)g^{-1}(0).
\end{equation}

This factorization holds generally, independent of the the specific form of the confinement potential. Also, the function ${\bf n}(\epsilon t)$ is assumed to be a general time-dependent unit vector. The factorization implies that if the confinement is strong enough, then $U_{2d}^{-1}H_\xi U_{2d}\approx0$ and $U_\xi\approx I$. It means that motion along the magnetic field direction and motion perpendicular to the magnetic field direction are described by two mutually commuting operators $U_{1d}$ and $U_{2d}$ in the strong confinement limit. This is the main result of the paper. Also, based on the expression of $H_\xi$ in terms of $\xi(t)$, corrections to the strong confinement approximation may be obtained once $U_{2d}$ is known.

Of relevance here is that $x_3-L$ and $U_\xi$ are operators, so the meaning of their closeness to the zero operator and the identity operator should be carefully analyzed. This is done in the following section where $U_\xi$ is studied based on the expression of $H_\xi$ which has just been derived.

We note that in the strong confinement limit $U_\xi\rightarrow I$, $U_{2d}$ still corresponds to a general time evolution in which the variation of ${\bf n}(\epsilon t)$ can be non-adiabatic. Only in the further limit $1/\epsilon\gg m/eB$, does it correspond to the adiabatic time evolution. The difference between the strong confinement condition and the adiabatic condition is an interesting point that is further examined in the following section.

The main step in the factorization of $U_1$ is the gauge transformation $|\psi_1(t)\rangle=g(\epsilon t)|\psi_0(t)\rangle$, under which
we switch from $U_1$ to $U_0$ through the relation $U_1=g(\epsilon t)U_0g^{-1}(0)$. In the gauge corresponding to $U_0$, the factorization is
the easiest to derive: $U_0=U_{1d}(t)U_{2d}(t)U_\xi(t)$. In the gauge corresponding to $U_1$, however, it is more natural to write the result as
\begin{equation}
U_1(t)=W_{1d}(t)W_{2d}(t)W_\xi(t)g(\epsilon t)g^{-1}(0),
\end{equation}
where $W_{1d}=g(\epsilon t)U_{1d}(t)g^{-1}(\epsilon t)$, and so on, still having the same physical contents and $g(\epsilon t)g^{-1}(0)$ serves to update the gauge of an initial state at $t=0$ to its gauge at time $t$.

\section{$U_\xi$ and the strong confinement limit}
The operator $\xi(t)=U_{1d}^{-1}(t)(x_3-L)U_{1d}(t)$ (or, for the same purpose, $\xi_0=x_3-L$) is naturally associated with the strength of the confining potential. From the physical point of view, when the strength of the confinement potential increases, the energy gaps between different energy levels of the Hamiltonian $H_{1d}=\frac{1}{2m}p_3^2+V(x_3-L)$ also increase. For a given eigenstate $|\Phi_n\rangle$ of $H_{1d}$, if $\langle\Phi_n|x_3-L|\Phi_n\rangle=0$, as is the case if the potential is symmetric with respect to the equilibrium position, then the smallness of $\xi(t)$ can be measured by the quantity
\begin{equation}
|\xi|=\sqrt{\langle\xi^2(t)\rangle}=\sqrt{\langle\Phi_n|(x_3-L)^2|\Phi_n\rangle}.
\end{equation}
If the minimum energy gap of $H_{1d}$ is $\Delta$, then the strong confinement limit can be understood as either $|\xi|\rightarrow0$ or $\Delta\rightarrow\infty$. The two simplest examples are the harmonic oscillator potential and the infinite potential well. These correspond to linear and nonlinear Heisenberg equations of motion, respectively. For both cases, we have $|\xi|^2=c(n)/\Delta$, where $c(n)$ depends only on the energy level $n$ of $H_{1d}$. Here we have three time scales, $\hbar/\Delta$, $1/\epsilon$, and $m/eB$, which provide the ratios needed for a perturbative treatment of $U_\xi$. The condition that the confinement is strong means that $\hbar/\Delta$ is much smaller than the other two time scales.

We note that the factorization of $U(t)$ here is not the same as $U=GDU_\epsilon$ in that it reduces to a simple form in the strong confinement
limit $|\xi|\rightarrow0$ rather than the limit $\epsilon\rightarrow0$ in which the rotation is infinitely slow. In the present problem, this factorization based on confinement strength is more general and more essential than a factorization based on rotation speed. This is because although $H_\xi$ is proportional to both $\xi(t)$ and $\epsilon$, the condition $|\xi|\rightarrow0$ corresponds to $U_\xi\rightarrow I$ but the condition $\epsilon\rightarrow0$ does not. To see this, let us examine the structure of $U^{-1}_{2d}H_\xi U_{2d}$, which determines $U_\xi$. From the expression of $H_\xi $, we see that the crucial term in $U^{-1}_{2d}H_\xi U_{2d}$ is,
\begin{equation}
U^{-1}_{2d}[-2\alpha_\mu(\epsilon t)(p_\mu-eA_\mu)-e\alpha_\mu(\epsilon t)A_\mu-m{\dot\alpha}_\mu(\epsilon t)x_\mu]U_{2d}\xi(t).
\end{equation}
This term causes the coupling between the 1D and 2D motions. It is proportional both to $\xi(t)$ and $\epsilon$. It might seem that either the strong confinement limit $\Delta\rightarrow\infty$ or the limit $\epsilon\rightarrow0$ is sufficient in suppressing the effects caused by this term. But this is not so. The limit $\epsilon\rightarrow0$ is not enough. This is because the time evolution happens in the interval $[0, 1/\epsilon]$, and terms of the order $\epsilon$ may accumulate to give finite contributions. In the case that these $\epsilon$ order terms are coupled to oscillating periodic functions with much shorter periods compared with $1/\epsilon$, they are averaged out and do not accumulate. This is the mechanism behind the proof of the
adiabatic theorem. Let us examine the term $-2\alpha_\mu U^{-1}_{2d}(\epsilon t)(p_\mu-eA_\mu)U_{2d}\xi(t)$ in the above expression. If $\epsilon\rightarrow0$, then $U^{-1}_{2d}(\epsilon t)(p_\mu-eA_\mu)U_{2d}$ should recover the behavior of $p_\mu-eA_\mu$ as the Heisenberg operator in the usual Landau problem (see
the next section), so it contains oscillating factors $\exp(\pm i\omega t)$. However, $\xi(t)$ is oscillating too and, in the simplest case of the harmonic oscillator, contains factors $\exp(\pm i\Delta t/\hbar)$. Then if $\omega$ matches $\Delta/\hbar$, such an averaging mechanism does not work. So $\epsilon\rightarrow0$ does not imply $U_\xi\rightarrow I$. The strong confinement condition $\Delta\rightarrow\infty$ is different because increasing $\Delta$ does not increase the duration of the time evolution. Thus, increasing $\Delta$ or, equivalently, decreasing $|\xi|$ against fixed values of $\epsilon$ and $\omega$ can make sure that the effects caused by $U^{-1}_{2d}H_\xi U_{2d}$ are small for an initial state with fixed quantum numbers. So we deduce that the strong confinement condition is a more succinct statement here in considering the limit $U_\xi\rightarrow I$. Furthermore, in addition to shrinking $|\xi|$, increasing $\Delta$ against fixed values of $\epsilon$ and $\omega$ automatically provides the fast periodic phase factors needed in averaging out the terms of the order $\epsilon$. This means that $U_\xi$ may in fact approach the identity operator faster than expected from the smallness of $|\xi|$ alone. This may be useful in a more refined analysis on the difference between $U_\xi$ and the identity operator.

Let us observe that if the potential is not close to the strong confinement limit, i.e., if the off-diagonal matrix elements of $x_3(t)-L$ are not small (i.e., the time scale $\hbar/\Delta$ is not sufficiently smaller than $1/\epsilon$ and $\omega^{-1}=m/(|e|B)$), then a perturbation approach may become ineffective and the study of $U_{\xi}(t)$ may require different methods. It is clear that $U_{\xi}(t)$ does not in general commute with $U_{1d}^{-1}(t)x_3U_{1d}(t)$ because of its time dependence. Then $U_{\xi}(t)$ necessarily contributes to the dynamics both for motion along the magnetic field and motion perpendicular to it, with $U_{1d}$ and $U_{2d}$ not necessarily the major contributions. Also, $U_{\xi}(t)$ is generated by the time-dependent Hamiltonian $U_{2d}^{-1}H_\xi U_{2d}$ that is neither linear nor quadratic, because $\xi(t)$ is in general a nonlinear function of $x_3$ and $p_3$. So the operator $U_{\xi}(t)$ contains the major complexity of the problem if the confinement is not strong. Nevertheless, $U_\xi$ is unique in the sense that it has concentrated the major complexities of the problem and that it approaches the identity operator in the strong confinement limit. For the general situation away from such a limit, perhaps this operator may contain useful structural information about the time evolution that can be extracted by methods beyond perturbation theory. One sees that $U_{2d}$ dictates a linear time evolution of the variables $p_\mu-eA_\mu$ and $x_\mu$ ($\mu=1,2$), so that the Hamiltonian $U_{2d}^{-1}H_\xi U_{2d}$
represents a simple and nontrivial example of time-dependent coupling between linear and nonlinear motions.

\section{$U_{2d}$ and non-adiabatic effects}
Under the strong confinement limit, $U_\xi=I$, the time evolution becomes
\begin{equation}
U(t)=R(\epsilon t)g(\epsilon t)U_{1d}(t)U_{2d}(t)g^{-1}(0).
\end{equation}
It is clear that in this limit, motion perpendicular to the magnetic field direction is determined by $U_{2d}(t)$.
Although $\epsilon$ does not have to be small compared with $\omega$, our focus here is to study $U_{2d}(t)$ by a perturbation method
which is most effective for small $\epsilon$.

Unlike $H_\xi$, the Hamiltonian $H_{2d}$ is quadratic and corresponds to a linear system. For this kind of systems, there has been extensive work focusing on the construction of quadratic invariants and the quantum propagator (See, for instance, \cite{lewis, dodonov, combescure, fiore}). However, due to the existence of the term $-\frac{m}{2}[\alpha_1(\epsilon t)x_1+\alpha_2(\epsilon t)x_2]^2$ in $H_{2d}$, the Heisenberg equations of motion are not solvable by quadrature for a general time-dependent function ${\bf n}(\epsilon t)$ \cite{bryant}. So in the present situation the quadratic invariants and the propagator cannot be constructed explicitly.

Our approach here is to factorize from $U_{2d}$ a major contribution in terms of an operator that is explicitly constructed. Effects due to the existence of small terms including $-\frac{m}{2}[\alpha_1(\epsilon t)x_1+\alpha_2(\epsilon t)x_2]^2$ are then studied by a perturbation method. In this way, $U_{2d}$ can be studied explicitly to any order in $\epsilon$. This approach also directly gives the Berry phase information contained in $U_{2d}$.

\subsection{Factorization of $U_{2d}$ and the adiabatic limit}
The Hamiltonian $H_{2d}$ is of the form
\begin{equation}
H_{2d}=\frac{1}{2m}[(p_1-eA_1)^2+(p_2-eA_2)^2]-eL\alpha_\mu(\epsilon t)A_\mu+S(\epsilon^2),
\end{equation}
where $S(\epsilon^2)$ contains terms that are of the order $\epsilon^2$:
\begin{equation}
S(\epsilon^2)=mL{\dot \alpha}_\mu(\epsilon t)x_\mu-\frac{m}{2}[\alpha_1(\epsilon t)x_1+\alpha_2(\epsilon t)x_2]^2.
\end{equation}
Since the system is linear, effects caused by $S(\epsilon^2)$ for the interval $[0, 1/\epsilon]$ is at most of the order $\epsilon$, which goes to zero if
$\epsilon\rightarrow 0$. However, the term $-eL\alpha_\mu(\epsilon t)A_\mu$, which is of the order $\epsilon$, may accumulate over $[0, 1/\epsilon]$ to have a finite effect. Denote
\begin{equation}
\tilde{H}_{2d}=\frac{1}{2m}[(p_1-eA_1)^2+(p_2-eA_2)^2]-eL\alpha_\mu(\epsilon t)A_\mu,
\end{equation}
\begin{equation}
\tilde{U}_{2d}(t)=\mathrm{T}\exp\{-i\int_0^t\tilde{H}_{2d}(\tau)d\tau\}.
\end{equation}
Then we have
\begin{equation}
U_{2d}=\tilde{U}_{2d}U_\epsilon
\end{equation}
where
\begin{equation}
U_\epsilon=\mathrm{T}\exp\{-i\int_0^t\tilde{U}_{2d}^{-1}S(\epsilon^2)\tilde{U}_{2d}d\tau\}.
\end{equation}
We first state a result about $\tilde{U}_{2d}(t)$.
Following Brown and Zak \cite{brown, zak1, zak2}, we let
\begin{equation}
\pi_\mu=p_\mu-eA_\mu,~~\eta_\mu=p_\mu+eA_\mu.
\end{equation}
They satisfy
\begin{equation}
[\pi_\mu, \eta_{\nu}]=0,\ \ \  [\eta_1, \eta_{2}]=-ieB,\ \ \
[\pi_1, \pi_2]=ieB.
\end{equation}
The operators $\eta_\mu,~\mu=1,2$ generate magnetic translations: in addition
to implementing
\begin{equation}
\exp(i\eta_1d_1+i\eta_2d_2)x_\mu\exp(-i\eta_1d_1-i\eta_2d_2)=x_\mu+d_\mu,
\end{equation}
just like ordinary translation operators, they commute with $\pi_\mu$. Assume $e<0$, say it is the charge of the electron. Then the cyclotron frequency is $\omega=-eB/m$.
Using $\pi_\mu$ and $\eta_\mu$, one may define
\begin{equation}
a=\frac{1}{\sqrt{-2eB}}(\pi_1-i\pi_2), ~~b=\frac{1}{\sqrt{-2eB}}(\eta_1+i\eta_2).
\end{equation}
Then $a$ and $b$ satisfy
\begin{equation}
[a, a^{\dagger}]=1, ~~[b, b^{\dagger}]=1.
\end{equation}
The usual Landau Hamiltonian can be written as
\begin{equation}
H_B=\frac{1}{2m}(\pi_1^2+\pi_2^2)=\hbar\omega(aa^\dagger+1/2).
\end{equation}
In \cite{chee1}, we studied the Landau problem with a time-dependent, spatially uniform electric field. The result in \cite{chee1} directly applies here
because $-eL\alpha_\mu(\epsilon t)A_\mu$ is the same as a linear electric field potential. We have
\begin{equation}
\tilde{U}_{2d}=M(\epsilon t)e^{-iH_Bt}{\tilde U}_\epsilon(t),
\end{equation}
where $M(\epsilon t)$ is a path-ordered magnetic translation:
\begin{equation}
M(\epsilon t)=\exp[-i\eta_1d_1(\epsilon t)-i\eta_2d_2(\epsilon t)]e^{i\beta(\epsilon t)},
\end{equation}
with $d_\mu(\epsilon t)$ and $\beta(\epsilon t)$ given by
\begin{equation}
d_\mu(\epsilon t)=-\frac{L}{2}\int_{0}^{t}\alpha_\mu(\epsilon\tau)d\tau=-\frac{L}{2}\int_{{\bf n}(0)}^{{\bf n}(\epsilon t)}{\bf e}_{\mu}\cdot d{\bf n},
\end{equation}
\begin{equation}
\beta(\epsilon t)=-eBS_d=-eB\big(\int_{0}^{t}d_1(\epsilon\tau)\big(-\frac{L}{2}\alpha_2(\epsilon\tau)\big)d\tau-\frac{1}{2}d_1(\epsilon t)d_2(\epsilon t)\big).
\end{equation}
We see that $S_d$ is the area enclosed by the path traversed by $(d_1(\epsilon t), d_1(\epsilon t))$ in the $(d_1, d_2)$ plane and the straight line pointing from $(d_1(\epsilon t), d_1(\epsilon t))$ to $(0, 0)$. Furthermore, it is clear that $(d_1(\epsilon t), d_1(\epsilon t))$ is fixed by the path of ${\bf n}(\epsilon t)$ on the 2-sphere, so $M(\epsilon t)$ is determined by the path of ${\bf n}(\epsilon t)$ on the 2-sphere. Therefore, $M(\epsilon t)$ is a geometric operator just as $R(\epsilon t)$ is. It corresponds to physically displacing a wave packet in the rotating frame through a path determined by $(d_1(\epsilon t), d_1(\epsilon t))$. The total length of the path is $\frac{L}{2}$ times the length of the path of ${\bf n}(\epsilon t)$ on the unit 2-sphere. The phase factor $e^{i\beta}$ is quantum mechanical in nature and arises from the fact that the magnetic translation group is noncommutative. The operator $J(t)$ is fixed by another path-ordered operator generated by $\pi_1$ and $\pi_2$, or equivalently, by $a$ and $a^\dagger$:
\begin{equation}
{\tilde U}_\epsilon(t)=e^{\delta(t)a-\delta^{\ast}(t)a^{\dagger}}e^{i\gamma(t)},
\end{equation}
where
\begin{equation}
\delta(t)=\frac{L}{4l_B}\int_0^t\alpha(\epsilon\tau)e^{-i\omega\tau}d\tau,
\end{equation}
\begin{equation}
\gamma(t)=i\int_{0}^{t}(\delta^{\ast}\dot{\delta}d\tau-\delta\dot{\delta}^\ast d\tau)=4S_\delta.
\end{equation}
and
\begin{equation}
l_B=1/\sqrt{-2eB}, ~~ \alpha=\alpha_1+i\alpha_2.
\end{equation}
Here $S_\delta$ is the area enclosed by the path traversed by $\delta(t)$ in the
complex $\delta$-plane and the straight line connecting the end and
initial points of the path. But unlike $d_\mu$, $\delta$ is not determined by the path of ${\bf n}(\epsilon t)$ alone. It also depends on
how the path is traversed in time. Moreover, because of the oscillating
factor $e^{-i\omega\tau}$ in the integral, the integral goes to zero as $\delta/\omega$. So $\delta$ should be considered as of the order $\epsilon$, and $S_\delta$ is of the order $\epsilon^2$. Therefore, $\tilde{U}_\epsilon(t)\rightarrow I$ when $\epsilon\rightarrow0$. In conclusion, $U_{2d}$ has a general expression£º
\begin{equation}
U_{2d}(t)=M(\epsilon t)e^{-iH_Bt}{\tilde U}_\epsilon(t)U_\epsilon(t),
\end{equation}
where $\tilde{U}_\epsilon(t)\rightarrow I$, $U_\epsilon(t)\rightarrow I$ in the limit $\epsilon\rightarrow0$.

From this we see that the adiabatic limit, which assumes $\epsilon\rightarrow0$ in addition to $U_\xi\rightarrow I$, implies that
\begin{equation}
U(t)=R(\epsilon t)M(\epsilon t)D(t),
\end{equation}
where $D(t)=e^{-iH_Bt}U_{1d}$

Under the assumption of the harmonic oscillator potential $V=\frac{1}{2}k\big({\bf r}\cdot{\bf n}(\epsilon t)-L\big)^2$, this limit was derived before by analyzing the limiting behavior of the solution of the Heisenberg equations of motion in the rotating frame. However, the gauge transformation $g(\epsilon)$, which is a crucial ingredient in deriving the general factorization of $U(t)$, was overlooked because $g(\epsilon)\rightarrow I$ when $\epsilon\rightarrow0$.

In the adiabatic limit and for a cyclic variation of ${\bf n}(\epsilon t)$ with ${\bf n}(\epsilon T_1)={\bf n}(1)={\bf n}(0)$, the Berry phase factor is essentially $R(1)M(1)$. For different paths of ${\bf n}(\epsilon t)$, these $R(1)M(1)$'s form a non-Abelian group, so the Berry phase factor is non-Abelian \cite{wil}

The fundamental discovery of Berry \cite{berry} (interpreted by Simon \cite {simon}) is that the holonomy group implemented by the Schr\"{o}dinger time evolution is not trivial. For the spin case, the $U(1)$ symmetry of the instantaneous Hamiltonian gives rise to the classical solid angle geometric phase\cite{berry}. Here, the symmetry group is much larger \cite{appelli}. It only fulfills the imagination that this larger symmetry is partly realized physically dictated by the Schr\"{o}dinger equation through the magnetic translation.

\subsection{Non-adiabatic effects}
Among the consequences of the factorization of $U(t)$, the operators $U_\xi$ and $\tilde{U}_\epsilon U_\epsilon$ can be used in calculating corrections to the limiting cases. Here we study non-adiabatic effects under the strong confinement assumption that $U_\xi\rightarrow I$. Then all non-adiabatic effects are in $\tilde{U}_\epsilon U_\epsilon$. Since $\tilde{U}_\epsilon$ is explicitly known, our focus here is the study of $U_\epsilon$.

Because $\tilde{U}_{2d}^{-1}S(\epsilon^2)\tilde{U}_{2d}$ is of the order $\epsilon^2$, $U_\epsilon-I$ is at most of the order $\epsilon$
for $t\in[0, 1/\epsilon]$. We have
\begin{equation}
U_\epsilon=I-i\int_0^t\tilde{U}_{2d}^{-1}S(\epsilon^2)\tilde{U}_{2d}d\tau+O(\epsilon^2).
\end{equation}
Furthermore, because $\tilde{U}_\epsilon=I+O(\epsilon)$, we have
\begin{equation}
U_\epsilon=I-i\int_0^te^{iH_B\tau}M^{-1}(\epsilon\tau)S(\epsilon^2)M(\epsilon\tau)e^{-iH_B\tau}d\tau+O(\epsilon^2).
\end{equation}
Since $x_\mu$ in $S(\epsilon^2)$ can be expressed in terms of $a, a^\dagger, b$, and $b^\dagger$, the integral can be calculated by using the relation $M(\epsilon t)^{-1}x_\mu M(\epsilon t)=x_\mu+d_\mu$ , $U^{-1}_B(t)bU_B(t)=b$, and $U^{-1}_B(t)a U_B(t)=ae^{-i\omega t}$. The result is
\begin{equation}
U_\epsilon=(1-ic_1)I+c_2b-c_2^\ast b^\dagger+c_3bb-c_3^\ast b^\dagger b^\dagger+c_4b^\dagger b+c_5a^\dagger a + O(\epsilon^2),
\end{equation}
where
\begin{equation}
c_1=\int_{0}^{t}[mL{\dot\alpha}_\mu d_\mu-\frac{m}{2}(\alpha_1d_1+\alpha_2d_2)^2-\dot{\bf n}^2ml_B^2]d\tau,
\end{equation}
\begin{equation}
c_2=\int_{0}^{t}\frac{1}{2}ml_B[L\dot{\alpha}^\ast-(\alpha_1d_1+\alpha_2d_2)\alpha^\ast]d\tau,
\end{equation}
\begin{equation}
c_3=-i\int_{0}^{t}\frac{1}{2}m{l_B}^2{\alpha^\ast}^2d\tau,~~~c_4=c_5=i\int_{0}^{t}m{l_B}^2\dot{\bf n}^2d\tau.
\end{equation}
In the above, the variables $\alpha_\mu$, $\alpha^\ast$, $d_\mu$ and $\dot{\bf n}$ under the integral sign are functions of $\tau$.
In deriving this result, terms proportional to integrals such as $\int_{0}^{t}\alpha^2(\epsilon\tau)e^{-2i\omega\tau}d\tau$ should be seen as of order $\epsilon^2$ because of the averaging effect of the oscillating factors. They are included in $O(\epsilon^2)$. The integrals in $c_1$ to $c_5$ should be
seen as of the order $\epsilon$ because effects of the order $\epsilon^2$ in these expressions accumulate over the time interval $[0, 1/\epsilon]$ and give
effects of the order $\epsilon$.

From the expression of $U_\epsilon(t)$, we see that the term $S(\epsilon^2)$ does not cause inter Landau level transitions up to the order $\epsilon$. Instead, the $c_1$ term represents a common phase correction for all wave functions, while
$c_5a^\dagger a$ represents a phase correction that depends on the Landau energy level. The other terms represent translations ($c_2$ and $c_2^\ast$) and distortions ($c_3, c_3^\ast$ and $c_4$) of a wave packet without changing its energy, which are corrections to the non-Abelian part of the geometric phase. The inter-Landau level transitions caused by the term $-eL\alpha_\mu(\epsilon t)A_\mu$ as represented by ${\tilde U}_\epsilon$ is equivalent to the influence of a uniform electric field which has been analyzed in detail in \cite{chee1}.

\section{Conclusions}
In this paper we have studied in detail the Landau electron problem in a rotating magnetic field and a rotating potential.
The Landau energy levels of the instantaneous Hamiltonian are infinitely degenerate and the potential is of a general form.
The key result of the paper is a natural factorization of the time evolution operator by
making use of the rotation operator technique and by a careful analysis of the structure of the Hamiltonian in the rotating frame.
The factorization is from a different perspective than the one suggested by the quantum adiabatic theorem and includes the latter as
a special case. The detailed structure regarding the coupling between motion in the magnetic field direction and motion perpendicular to
it is carefully analyzed. It is pointed out that it represents a simple and nontrivial example of the coupling between linear and nonlinear motions.

In the strong confinement limit, the problem is reduced to a two-dimensional one. We presented a different method
so that effects due to the quadratic potential can be analyzed perturbatively.
The time evolution operator for this two-dimensional motion is completely determined to include all $\epsilon$ order terms.

Since the Landau electron problem and the rotating magnetic field are often encountered in different areas in physics, the method and results presented here
may find their use in the study of various physical situations associated with applications.

\end{document}